# Non-monotonic size dependence of the elastic modulus of nanocrystalline ZnO embedded in a nanocrystalline silver matrix

Vinod Panchal, Shankar Ghosh[1], Smita Gohil, Nilesh kulkarni, Pushan Ayyub

Department of Condensed Matter Physics and Materials Science,
Tata Institute of Fundamental Research,
1 Homi Bhabha Road,
Mumbai 400005, India

## Abstract

We present the first high pressure Raman study on nanocrystalline ZnO films with different average crystallite sizes. The problem of low Raman signals from nano sized particles was overcome by forming a nanocomposite of Ag and ZnO nanoparticles. The presence of the nanodispersed Ag particles leads to a substantial surface enhancement of the Raman signal from ZnO. We find that the elastic modulus of nanocrystalline ZnO shows a non-monotonic dependence on the crystallite size. We suggest that the non-monotonicity arises from an interplay between the elastic properties of the individual grains and the intergranular region.

---

[1] Corresponding author emal : sghosh@tifr.res.in

**Introduction**

A decrease in the crystallite size is usually found to produce an increase in the bulk modulus[1-2], though certain exceptions are known[3]. The observed elevation of the bulk modulus with decreasing size has been rationalized on the basis of the surface energy of nanoparticles being greater than that of the bulk[4]. Though, isolated ZnO nanowires are known to show a significant increase in the elastic modulus with lowering of the wire diameter,[5] we do not know of any study of the elastic properties of nanocrystalline ZnO. This paper presents a study of the effect of crystallite size on the elastic modulus of nanocrystalline ZnO thin films, obtained from high pressure Raman spectroscopy. The nanocrystalline ZnO thin films are essentially random mosaics of crystallites (see Fig.1 (a)). We show that the elastic modulus of such a system is a non-monotonic function of the crystallite size, and attribute this behavior to a competition between the elastic properties of individual nano-domains and those of the intercrystalline region. The application of stress on a system results in the development of strain, to which phonon frequencies are extremely sensitive. High pressure Raman spectroscopy is therefore a very powerful technique for studying elastic properties, but often suffers from extremely low signal intensities, especially for nanocrystalline samples. We succeeded in circumventing this difficulty by using a dilute, random dispersion of nanocrystalline Ag in a ZnO matrix (also nanocrystalline), which leads to substantially intensified signals due to the surface enhancement mechanism[6].

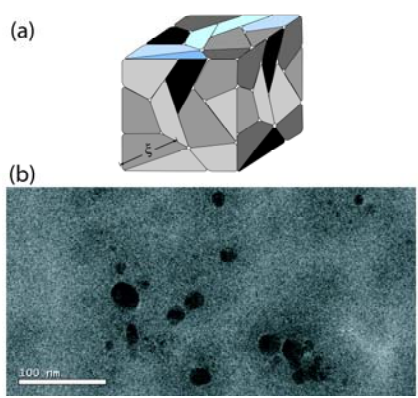

**Figure 1. (a) Schematic representation of the random mosaic structure of nanocrystalline domains with an average crystallite size $\xi$. (b) TEM of the Ag nanoparticles (darker contrast) dispersed in the nanocrystalline ZnO matrix.**

**Experimental Details**

Magnetron sputtering at relatively high ambient pressure and low substrate temperature is known to produce nanocrystalline thin films in which the grain size can be controlled by proper choice of process parameters[7]. A ZnO target (50 mm diameter, 3mm thick) was prepared by pressing ZnO powder (99.99 %) at 0.3 Gpa and annealing at 400°C in Ar. Two small silver pieces were placed on the horizontally mounted ZnO target. Nanocomposite films of Ag and ZnO (nc-ZnO/Ag) with different average particle sizes was deposited on glass substrates (at 300K) by sputtering from the ZnO/Ag composite target. RF sputtering was carried out typically at 100-150 W power under 5-200 mTorr Ar pressure. The as-prepared film, with a black metallic luster, was peeled off the glass substrate and used for further studies. Transmission electron microscope (TEM) observations indicate that the Ag nanoparticles were dilutely dispersed in the nanocrystalline ZnO matrix. The darker regions in Fig.1 (b) are the Ag nanoparticles while the grey background is nanocrystalline ZnO. The coherently diffracting crystallite size ($\xi$) was estimated from an analysis of the x-ray diffraction line broadening. X-ray line profile analysis was performed using the GSAS software[8], from which we obtained the full width at half maximum (FWHM), $\beta$, for a number of diffraction peaks. The crystallite sizes were then obtained from the FWHM using the Williamson Hall method[9]. The refined x-ray diffraction patterns and corresponding williamson Hall plots for a few representative samples are shown in Fig. 2.

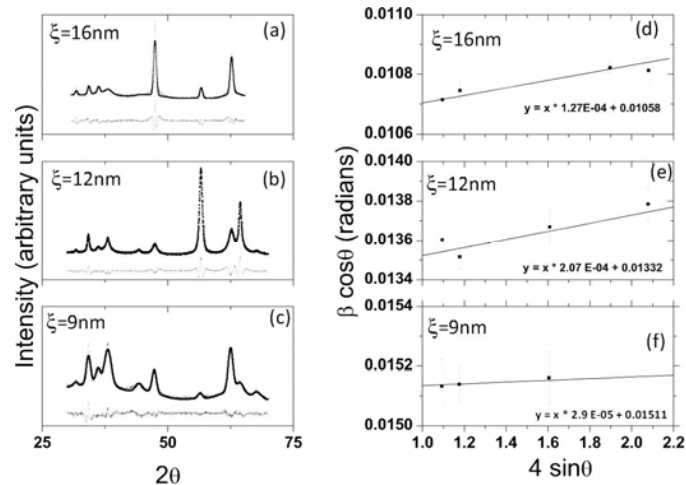

**Figure 2. (a), (b), (c) are the x-ray profile fits using GSAS and (d), (e) and (f) are the corresponding Williamson-Hall plot for $\xi$ = 16, 12 and 9nm respectively. In the Williamson-hall plot $\beta \cos \theta$ is plotted against $4 \sin \theta$. The solid line is the linear fit.**

Raman scattering measurements were performed in the back scattering geometry using a Jobin Yvon T64000 Raman spectrometer equipped with an inverted microscope. The excitation source was the 514.5 nm line of an Ar+ laser operated at a power of about 100 mW, measured at the sample. Figure 3 (a) shows the Raman spectrum of bulk ZnO powder. The spectrum shows the $A_1$(TO) mode at 381 cm$^{-1}$ $E_1$(TO) at 412 cm$^{-1}$, $E_{2h}$ at 438 cm$^{-1}$ and the $E_1$(LO) at 580 cm$^{-1}$. The feature at 537 cm$^{-1}$ has been attributed to a combinational mode (CM)[10].

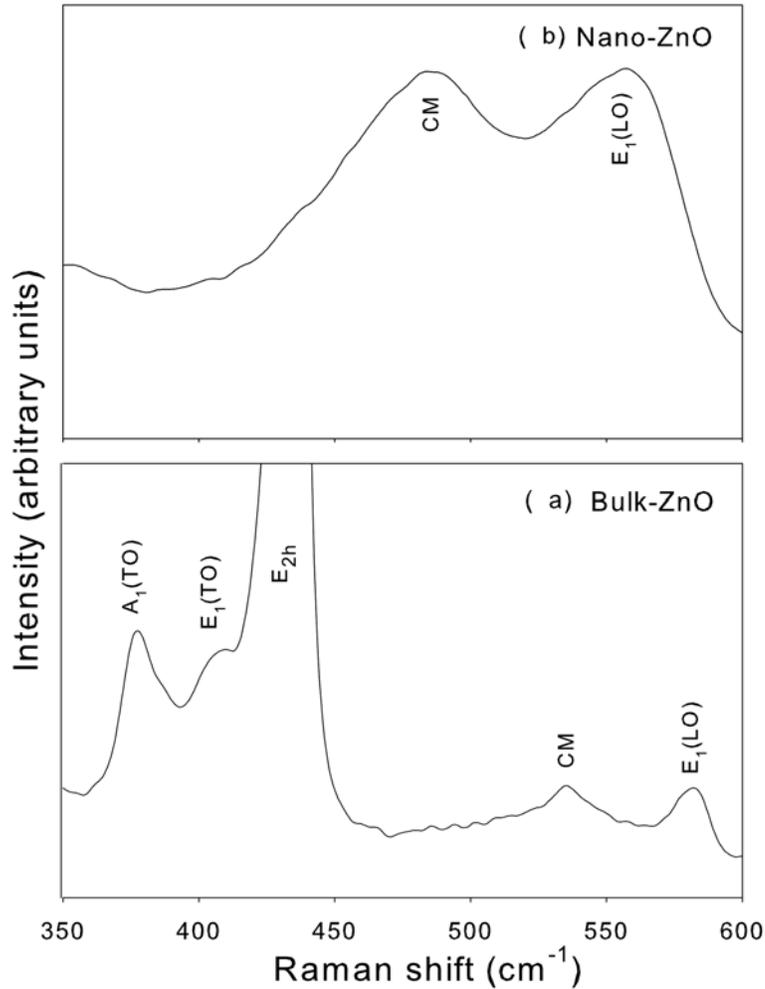

**Figure 3. Room temperature and ambient pressure Raman spectra of (a) bulk ZnO powder (B-ZnO) and (b) nanocomposite ZnO/Ag sample with an average particle size of 16nm. The mode assignments are shown in the figure.**

Figure 3(b) shows the Raman spectrum of the nc-ZnO/Ag composite. The CM and the $E_1$(LO) mode for the ZnO nanoparticles dispersed in the Ag matrix are red shifted to 484 cm$^{-1}$ and 560 cm$^{-1}$, respectively. The observed red shift of the Raman modes in nc-ZnO/Ag can be attributed to finite size effects that are known to lead to a breakdown of the wave vector selection rule and allow phonons away from the Brillouin zone center to also participate in the Raman scattering process. In addition, the tensile stress at the ZnO/Ag interface, due to a difference in their thermal expansion coefficients, may also contribute to the red shift. A key feature of the Raman spectra of nc-ZnO/Ag is the absence of the normally strongest $E_{2h}$ mode. This may be explained from the fact that the LO modes gets selectively enhanced under resonant conditions due to the breakdown of the wave-vector dependent scattering rules arising from the contributions to the scattering intensity from the three band electro-optic Raman tensor[11]. This suggests that the presence of Ag near the nanocrystalline ZnO generates a resonance-like condition[11].

High pressure experiments were performed in a gasketed Mao-Bell type diamond anvil cell (DAC) with a 4:1 mixture of methanol and ethanol as the pressure transmitting medium. A stainless steel gasket was pre-indented to 80 μm, and a 200 μm hole was mechanically drilled at the center. Before and after recording each Raman spectrum, the pressure was measured using the fluorescence emission of a small ruby piece placed close to the sample in the gasket. In these experiments, the maximum pressure applied was 10 GPa. We have restricted our experiments below 10 Gpa to avoid the known experimental issues of non hydrostatic nature of the pressure transmitting medium[12].

**Result and Discussion**

Figure 4 shows the pressure dependence of the Raman modes with increasing and decreasing pressure for representative nc-ZnO/Ag samples with ZnO crystallite sizes of 16nm, 12nm and 9nm. The $E_1$(LO) as well as the CM modes were found to increase monotonically with pressure. Bulk ZnO is known to undergo a crystallographic phase transition from the wurtzite to the cubic structure at 9 - 10 Gpa[13,14]. Since there are no Raman active phonons in the cubic phase, the wurzite to cubic transition is necessarily associated with a dramatic decrease in the Raman intensity[13]. However, we observe no such transition in nanocrystalline ZnO thin films up to 10 GPa. It is interesting to point out that a size-induced elevation of the structural transition pressure has been observed earlier[15].

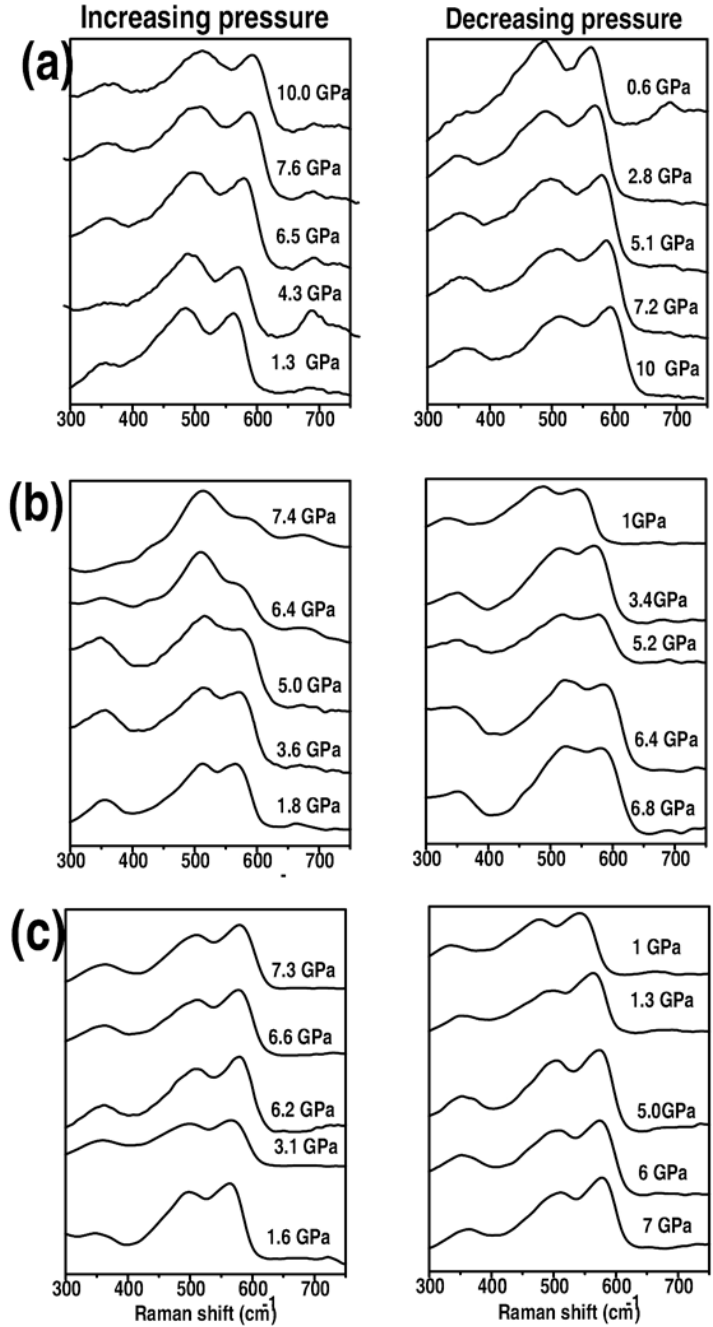

**Figure 4. The Raman Spectra of nc-ZnO/Ag samples recorded at different pressure. The left panel shows data recorded while increasing pressure while the right panel shows decreasing pressure data (a) $\xi$ = 16nm (b) $\xi$ = 11 nm and (c) $\xi$ = 9nm**

Figure 5 shows the variation of the mode frequencies of the $E_1$(LO) Raman mode as a function of increasing and decreasing pressure for three different nc-ZnO/Ag samples with mean sizes of 9 nm, 12 nm and 16 nm respectively. Clearly, the mode frequencies increase linearly with pressure for all crystallite sizes. The shift in the mode frequency is proportional to the strain and the applied pressure is the stress. ($\Delta\omega/\Delta p$) for the $E_1$(LO) mode was calculated from the Raman data to obtain the elastic modulus for each nanocomposite system. The elastic modulus: $G = \text{Stress}/\text{strain} \propto (\Delta\omega/\Delta P)^{-1}$

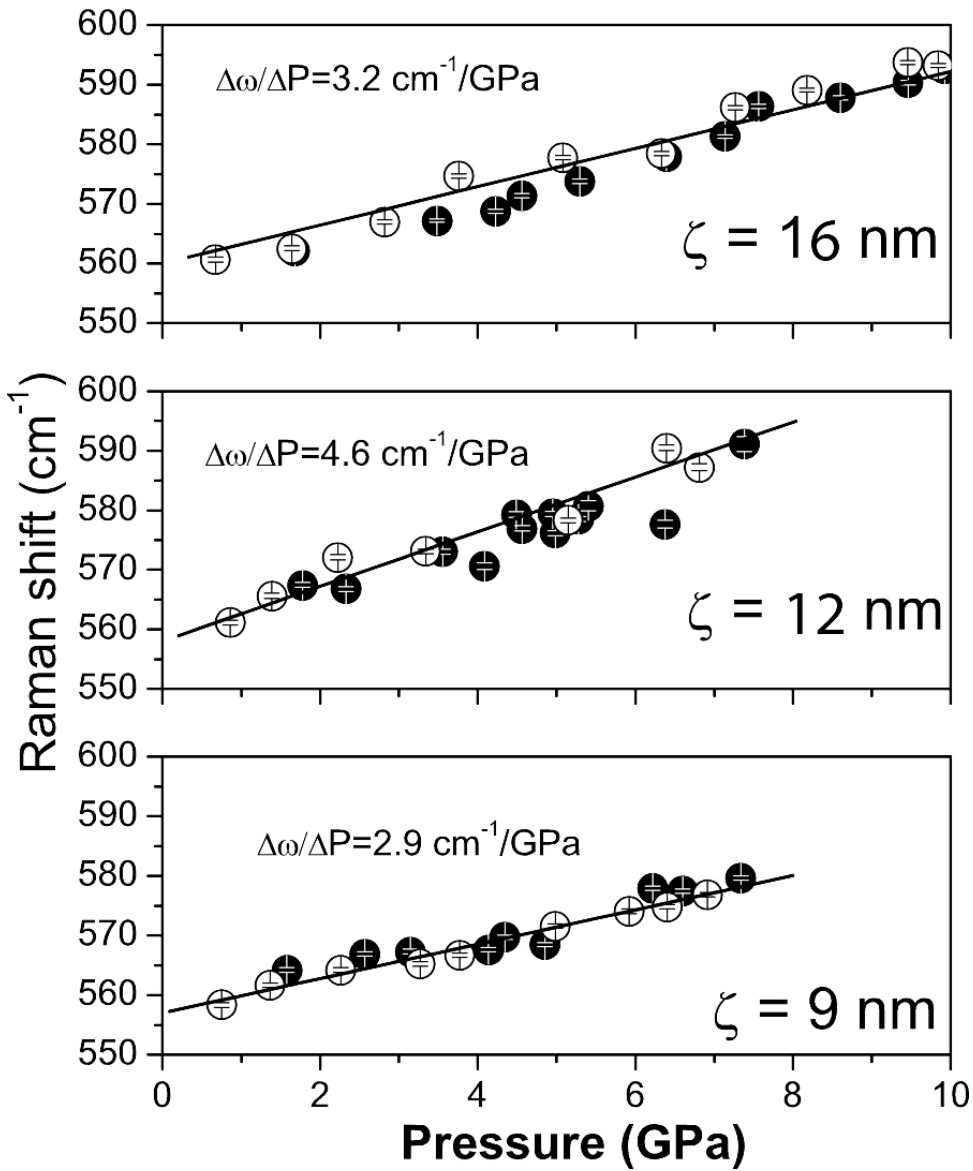

**Figure 5. Variation of the $E_1$(LO) Raman mode with pressure for nanocrystalline ZnO films with crystallite sizes of 16nm, 12nm and 9nm. The respective crystallite sizes (ξ) and $\Delta\omega/\Delta p$ are marked in the figure. The closed circles represent increasing pressure runs while the open circles represent decreasing pressure runs.**

In the present experiment the pressure transmitting liquid does not reach the surface of each ZnO nanocrystal. Such a non-hydrostatic pressure condition would allow volume conserving as well as shape conserving deformations. The elastic energy (U) per unit volume associated with a general deformation is given by $U = \mu ( u_{ik} - \frac{1}{3}\delta_{ik}u_{ll}^2 ) + \frac{1}{2}K u_{ll}^2$ [16] where K is the *bulk modulus* and µ is the *shear modulus*. The bulk modulus (K) is associated only with hydrostatic compression which results in a change in volume but none in shape. Since, the pressure conditions are not hydrostatic in the present experiments, extraction of parameters such as the bulk modulus (K) from high pressure Raman experiments is technically not possible. We therefore study the variation of $\Delta P/\Delta\omega$ which is a measure of an *average* elastic modulus as a function of crystallite size ($\xi$), as shown in Figure 6. Interestingly, $(\Delta\omega/\Delta P)^{-1}$ or the *average elastic modulus* is not a monotonic function $\xi$: with a well-defined minimum at about 13 nm.

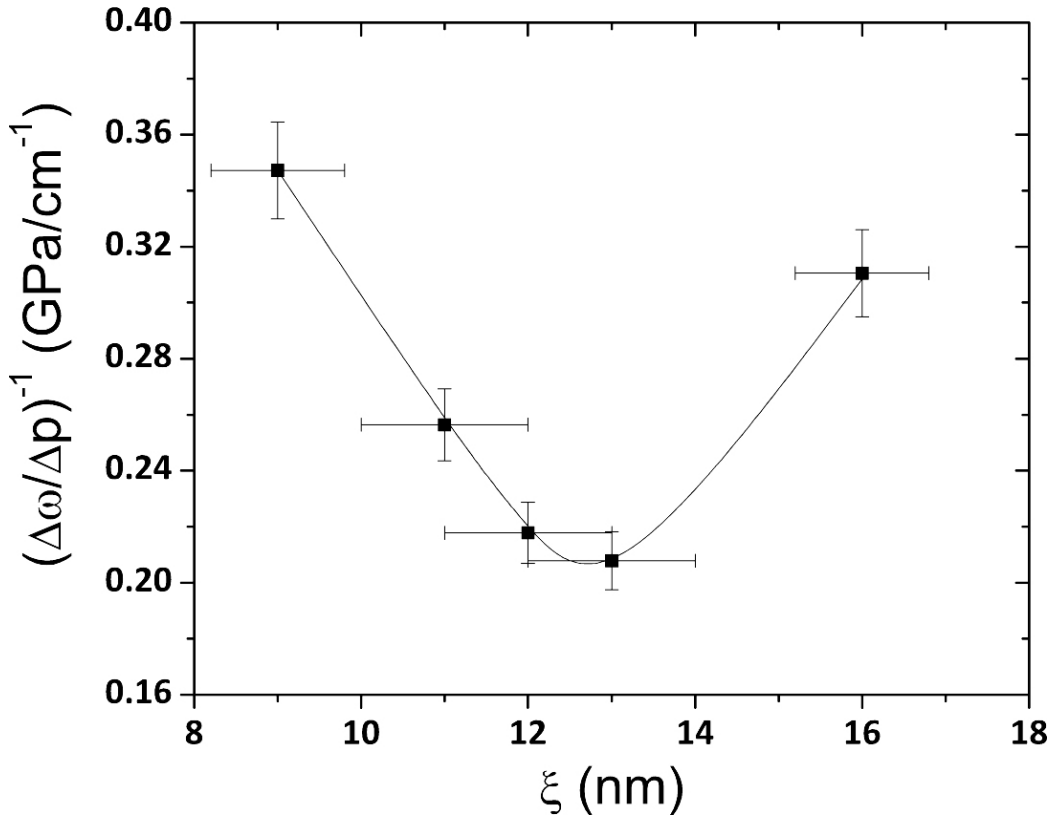

**Figure 6. Variation of $(\Delta\omega/\Delta P)^{-1}$ with size ($\xi$) of the ZnO nanocrystals. ($\xi$) was obtained from x-ray diffraction line broadening using the Williamson Hall method.**

To understand the elastic behavior of nc-ZnO/Ag, we model each nanocrystalline grain of size $\xi$ in terms of a (i) crystalline core and (ii) an intergranular region of average thickness L. The elastic modulus of the nanocomposite system is thus, $G=(1-p)E_c+pE_L$ where $p \propto L/\xi$ is the volume fraction of the intergranular region, $E_c$ is the elastic modulus of the core and $E_L$ is the elastic modulus of the intergranular region. All the individual nanocrystalline domains are assumed to have the same crystal structure, varying only in the crystallographic orientation. In the intergranular region It is believed that, in general, the crystal structure is amorphous like and $E_c > E_L$[17]. With a decrease in size, the volume fraction occupied by the intergranular region is expected to increase, which would result in the overall reduction of G. However at still smaller sizes the observed increase in G probably indicate an increase in $E_c$, which could arise from a build up of the residual compressive stress in the bulk[18].

**Conclusions**

In summary, we reiterate the salient features of this paper. (i) We have reported the first high pressure Raman study on nanocrystalline ZnO for a range of crystallite sizes from 9 nm to 16 nm. (ii) We have circumvented the problem of low Raman signals from from nanosized particles by incorporating nanocrystalline Ag in the form of a nanocomposite struture. The presence of the nanodispersed Ag particles leads to a surface enhancement of the Raman signal from the host (nano-ZnO). (iii) We have shown that the average elastic modulus of nanocrystalline ZnO shows a non monotonic variation with crystallite size. We propose that the non-monotonic behavior is due to an interplay between the elastic properties of the individual grains and those of the intergranular region.